\documentclass[aps,prl,twocolumn,longbibliography,superscriptaddress]{revtex4-1}

\usepackage{amssymb}
\usepackage{amsbsy}
\usepackage{amsmath}
\usepackage{graphicx}
\usepackage{graphics}
\usepackage{setspace}
\usepackage{array}
\usepackage{color}
\usepackage{fontenc}
\usepackage{textcomp}
\usepackage{bm}
\usepackage{float}
\usepackage[bookmarks=false,linkcolor=blue,urlcolor=blue,colorlinks,citecolor=blue]{hyperref}

\DeclareMathOperator{\im}{Im}

\newcommand{\vex}[1]{\bm{\mathrm{#1}}}

\newcommand{\bsub}{\begin{subequations}}
\newcommand{\esub}{\end{subequations}}

\graphicspath{{../Figures/}}

\begin{document}
\title{Non-Hermitian Higher-Order Dirac Semimetals}
\author{Sayed Ali Akbar Ghorashi}\email{sghorashi@wm.edu}
\affiliation{Department of Physics, William $\&$ Mary, Williamsburg, Virginia 23187, USA}
\author{Tianhe Li}
\affiliation{Department of Physics and Institute for Condensed Matter Theory,  University of Illinois at Urbana-Champaign, IL 61801, USA}
\author{Masatoshi Sato}
\affiliation{Yukawa Institute for Theoretical Physics, Kyoto University, Kyoto 606-8502, Japan}
\author{Taylor L. Hughes}
\affiliation{Department of Physics and Institute for Condensed Matter Theory,  University of Illinois at Urbana-Champaign, IL 61801, USA}

\date{\today}

\newcommand{\be}{\begin{equation}}
\newcommand{\ee}{\end   {equation}}
\newcommand{\bea}{\begin{eqnarray}}
\newcommand{\eea}{\end{eqnarray}}
\newcommand{\h}{\hspace{0.30 cm}}
\newcommand{\vs}{\vspace{0.30 cm}}
\newcommand{\n}{\nonumber}

\begin{abstract}
In this article we study 3D non-Hermitian higher-order Dirac semimetals (NHHODSMs). Our focus is on  $C_4$-symmetric non-Hermitian systems where we investigate inversion ($\mathcal{I}$) or time-reversal ($\mathcal{T}$) symmetric models of NHHODSMs having real bulk spectra. We show that they exhibit the striking property that the bulk and surfaces are anti-PT and PT symmetric, respectively, and so belong to two different topological classes realizing a novel non-Hermitian topological phase which we call a \emph{hybrid-PT topological phases}. Interestingly, while the bulk spectrum is still fully real, we find that exceptional Fermi-rings (EFRs) appear connecting the two Dirac nodes on the surface. This provides a route to probe and utilize both the bulk Dirac physics and exceptional rings/points on equal footing. Moreover, particularly for $\mathcal{T}$-NHHODSMs, we also find real hinge-arcs connecting the surface EFRs. We show that this higher-order topology can be characterized using a biorthogonal real-space formula of the quadrupole moment. Furthermore, by applying Hermitian $C_4$-symmetric perturbations, we discover various novel phases, particularly: (i) an intrinsic $\mathcal{I}$-NHHODSM having hinge arcs and gapped surfaces, and (ii) a novel $\mathcal{T}$-symmetric skin-topological HODSM which possesses both topological and skin hinge modes. The interplay between non-Hermition and higher-order topology in this work paves the way toward uncovering  rich phenomena and hybrid functionality that can be readily realized in experiment.
\end{abstract}
\maketitle

\emph{Introduction}.--
Topological phases matter have been on the frontier of condensed matter and related areas of research over last two decades \cite{Chiu2016,reviewweyl}. Recently, two new research directions have garnered attention: higher-order topology and non-Hermitian topological phases. The hallmark of a $d$-dimensional $nth$-order topological phase is a gapped bulk harboring robust boundary states/features with co-dimension $p=d-n$\cite{Benalcazar2017-1,Benalcazar2017-2,Schindler2018-1,Schindler2018-2, Song2017, Langbehn2017, benalcazar2019,tianhe1,GHHRHOTSC,ghorashi2019vortex,GLHHOWSM,CAlugAru2018}. Higher order topological systems have been experimentally observed in various physical platforms \cite{Peterson2018,Noh2018,Serra-Garcia2018,Imhof2018,ni2019observation,xue2019acoustic}.
Along with this, non-Hermitian topological phases are inherently realized in non-equilibrium contexts, and show many interesting properties with broad applications ranging from photonics to ultracold atoms\cite{ReviewNHRMP,reviewNHUeda,zhou2018observation,cerjan2019experimental,Gao_2021}. Similar to Hermitian topological phases, the topological classification of non-Hermitian phases has been extensively explored \cite{SatoPRX2018,SatoPRX2019,NHclassLiuPRB2019,NHclassSatoPRL2019,NHclassZhouPRB2019}. Interestingly,  the notion of bulk-boundary correspondence for non-Hermitian systems has proven to be subtle, primarily due to the emergence of exceptional structures (ESs) (points/rings/disks), on which eigenstates and eigenengergies coalesce in complex energy space \cite{heiss2012physics,ReviewNHRMP,reviewNHUeda}, as well as the phenomenon of the non-Hermitian skin effect where there is an extensive localization of states on the boundaries\cite{BiorthogonalPRL2018,ReviewNHRMP,SatoskinPRL2020,Skinnonbloch,NHskinopen,alvarez2018topological,NHskinTorres,Xiao_2020,PhysRevLett.124.250402,zou2021observation,helbig2020generalized,Ghatak_2020,weidemann2020topological}.
\begin{figure}[t!]
    \centering
    \includegraphics[width=0.5\textwidth]{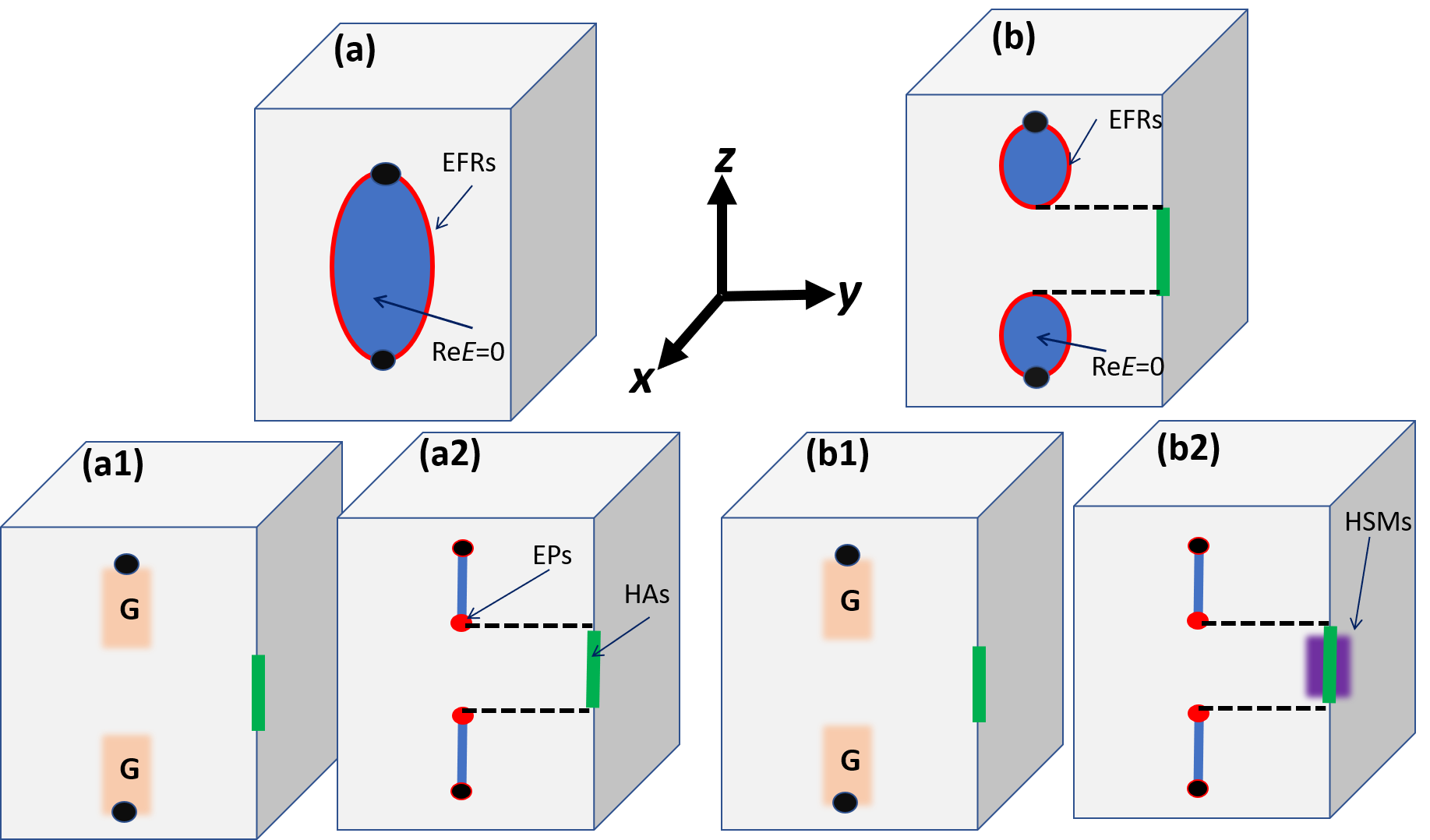}
    \caption{The schematic summary of results. (a)  $\mathcal{I}$-symmetric NHHODSMs can possess surface exceptional Fermi-rings (EFRs) connecting the projection of the real bulk Dirac nodes ( black dots) on the surface and no hinge arcs (HAs) while (b) $\mathcal{T}$-symmetric NHHODSMs can exhibit two patches of surface EFRs connected by HAs (green line on the hinge). (a1, b1) have turned on the $\alpha_1$ perturbation (a1,b1) for both models. In both cases the EFRs gap out and HAs appear.  (a2,b2) have $\alpha_2$ turned on for both models. This perturbation deforms the EFRs to exceptional points (EPs), [note that the projection of the real bulk Dirac nodes and two of EPs coincide] to form two exceptional Fermi-arcs on the surface which are connected by HAs on the hinges. (b2) Also possess hinge skin modes (HSMs) [shaded purple region] coexisting with HAs.}
    \label{fig:adpic}
\end{figure}

In this work, we study the interplay of higher-
order and non-Hermitian topological phases in the context of higher order topological semimetals. While previous studies have mostly explored two-dimensional and/or gapped phases \cite{Nori2ndorderPRL2019,EzawaHONHPRB2019,BiorthHOPRB2019,zhang2ndordersonic,hybridskintop}, the three-dimensional non-Hermitian semimetallic phases remain unexplored. We will begin by introducing non-Hermitian perturbations to a previously studied $C^z_4$-symmetric higher-order Dirac semimetal (HODSMs) \cite{Lin2017}. From this we will identify two novel phases: an $\mathcal{I}$-symmetric and a $\mathcal{T}$-symmetric non-Hermitian Dirac semimetal (NHDSM).  These phases feature real bulk spectra having point Dirac nodes, and a variety of  complex surface spectra. Strikingly, the bulk and surface in these phases possess anti-PT and PT symmetry respectively, and hence belong to two different classes of the non-Hermitian topological classification. This allows for new types of \emph{hybrid-PT non-Hermitian topological phases} where the bulk and surface ESs are protected by different symmetries. Both the $\mathcal{I}$ and $\mathcal{T}$-symmetric NHDSMs we study host exceptional Fermi ring (EFR) surface modes protected by symmetries of the non-Hermtian $\mathcal{P}$CI class. Additionally, the $\mathcal{T}$-symmetric phase hosts real hinge arcs (HAs), which can be characterized by a quantized bulk, biorthogonal quadrupole moment, and represents an unprecedented non-Hermitian higher order Dirac semimetal (NHHODSM) phase. Furthermore, by breaking the  symmetries that protect the surface EFRs while preserving the bulk $C^z_4$ symmetry, we unveil various novel phases including: (i) an  $\mathcal{I}$-NHHODSM having exceptional Fermi-arcs on the surface coexisting  with intrinsic hinge arcs having complex energies, (ii) a $\mathcal{T}$-symmetric skin-topological NHHODSM which shows both the hinge skin modes (HSMs) and complex HAs.

\emph{Models}.---To illustrate all of the phenomena mentioned above we can start with a parent Hermitian model of a HODSM introduced in \cite{Lin2017},
\begin{align}\label{hodsm1}
H_{HODSM}(\vex{k}) = \sum^{4}_{i=1} a_i(\vex{k})\Gamma_i,
\end{align}
where $a_1(\vex{k})=\sin(k_y),\, a_2(\vex{k})=\left(\gamma+\frac{1}{2}\cos k_z+\cos k_y\right),\, a_3(\vex{k})=\sin(k_x),\, a_4(\vex{k})=\left(\gamma +\frac{1}{2}\cos k_z + \cos k_x\right)$, and the $\{\Gamma_\alpha\}$ are direct products of Pauli matrices, $\sigma_i,\kappa_i$: $\Gamma_0=\sigma^3\kappa^0,\Gamma_i=-\sigma^2\kappa^i\,\textrm{for}\,i=1,2,3,$ and $\Gamma_4=\sigma^1\kappa^0.$ The parameter $\gamma$ represents the intra-cell coupling,
and we have set the amplitudes of the inter-cell couplings to $1$. This model preserves $C^z_4, $ mirror $\mathcal{M}_{x,y,z} $, inversion, $\mathcal{I}$, and time-reversal $\mathcal{T}=K$ \cite{note3} symmetries. We focus on the parameter regime $-0.5 < \gamma < -1.5$ where Eq. \eqref{hodsm1} generates a higher order Dirac semimetal spectrum exhibiting two gapless Dirac nodes in the bulk, gapped surfaces, and Fermi arcs connecting the nodes at the hinges. From this parent state we will investigate both inversion and time-reversal symmetric non-Hermitian perturbations and discuss some resulting topological phases.

\emph{$\mathcal{I}$-Model}.--- First we will consider the following non-Hermitian $\mathcal{I}$ symmetric model,
\begin{align}\label{I-NHHODSM}
    H^{\mathcal{I}}_{D} = H_{HODSM}(\vex{k}) + im_1\Gamma_0 a_0(\vex{k}),
\end{align}
where $a_0(\vex{k})=\left(\cos(k_x)-\cos(k_y)\right),$ and $m_1$ is a real constant. The $m_1$ term breaks $\mathcal{M}_{x,y}$ symmetries while keeping $\mathcal{I}$ and $C^z_4$. Interestingly, while $m_1\neq 0$ breaks $\mathcal{T}$ and hence $\mathcal{I}\mathcal{T}$, it preserves the anti-PT symmetry  $(\mathcal{I}\mathcal{T})H(\vex{k})(\mathcal{I}\mathcal{T})^{\dagger}=-H(\vex{k})$. Moreover, reciprocity $\mathcal{R}H(\vex{k})\mathcal{R}^{\dagger}=H^{T}(-\vex{k})$ and $\mathcal{I}\mathcal{R}$ are also preserved, where $\mathcal{R}=\mathbb{I}$. These symmetries, along with the chiral symmetry $\Gamma_0 H(\vex{k})\Gamma_0^{\dagger}=-H^{\dagger}(\vex{k})$, place  Eq.~\eqref{I-NHHODSM} in the class $\mathcal{P}$CII$^{\dagger}$ \cite{NHclassSatoPRL2019}.
Remarkably, the bulk spectrum remains fully real up to a critical value of $m_1^c$.
To see this we calculate the eigenvalues of Eq.~\eqref{I-NHHODSM}: $E(\vex{k})=\pm\sqrt{\sum^4_i a^2_i(\vex{k})-m_1^2a^2_0(\vex{k})}$.
The spectrum becomes complex at a given ${\bf{k}_0}$ only when $m_1 > m_1^c\equiv {\rm{min}}_{\bf{k}_0}\sum^4_i a^2_i(\vex{k}_0)/a^2_0(\vex{k}_0)$ (if $a^2_0(\vex{k}_0)=0$ this never happens). When the bulk energies become complex, 3D ESs form that project to four-fold degenerate exceptional rings in the $k_y-k_z$ and $k_x-k_z$ planes centered around $(k_x, k_z)=(\pi, 0)$ and ($k_y, k_z)=(\pi, 0)$, respectively (see \cite{sm}). Unlike many previous cases \cite{ReviewNHRMP} the ESs here emerge from the bulk away from the locations of the real Dirac nodes and not by transforming the gapless Dirac points to exceptional rings (or pair of EPs). Therefore, in this system the real Dirac nodes and the bulk ESs can coexist.
\begin{figure}[t!]
    \centering
    \includegraphics[width=0.4\textwidth]{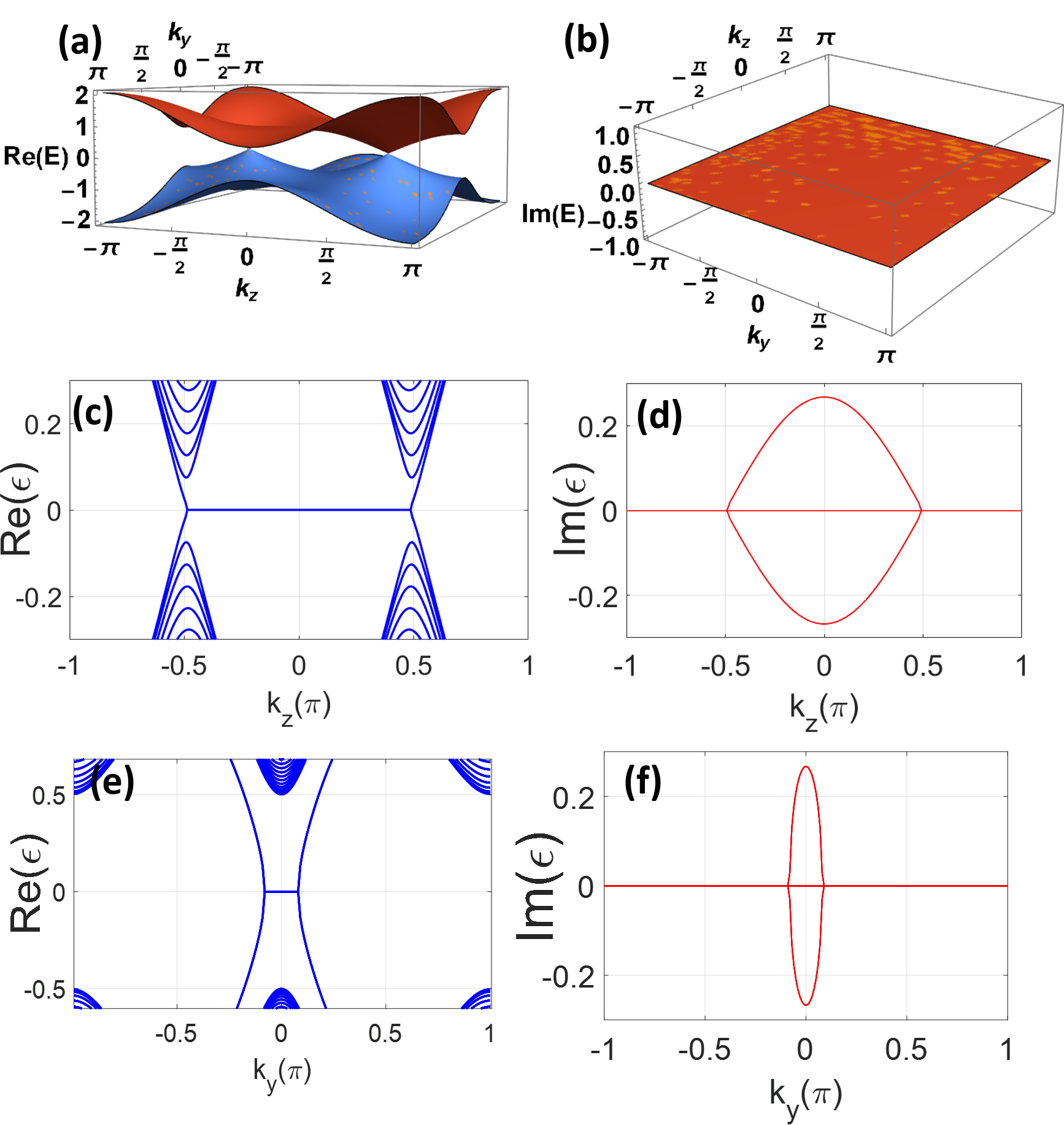}
    \caption{The spectrum of $H^{\mathcal{I}}$ with $\gamma=-1,\, m_1=-0.75\,(< m_1^c)$. (a,b) The real and imaginary parts of the bulk spectrum in the $k_y-k_z$-plane with periodic boundary conditions in all directions, (c,d) the real and imaginary spectrum for open boundaries conditions in the $x$-direction which depicts EFRs along the $k_z-axis,\,(k_y=0)$, (e,f) are similar to (c,d) except plotted along the $k_y-axis,\,(k_z=0)$.}
    \label{fig:I-NHHODSM}
\end{figure}
\begin{figure*}[htb!]
    \centering
    \includegraphics[width=0.9\textwidth]{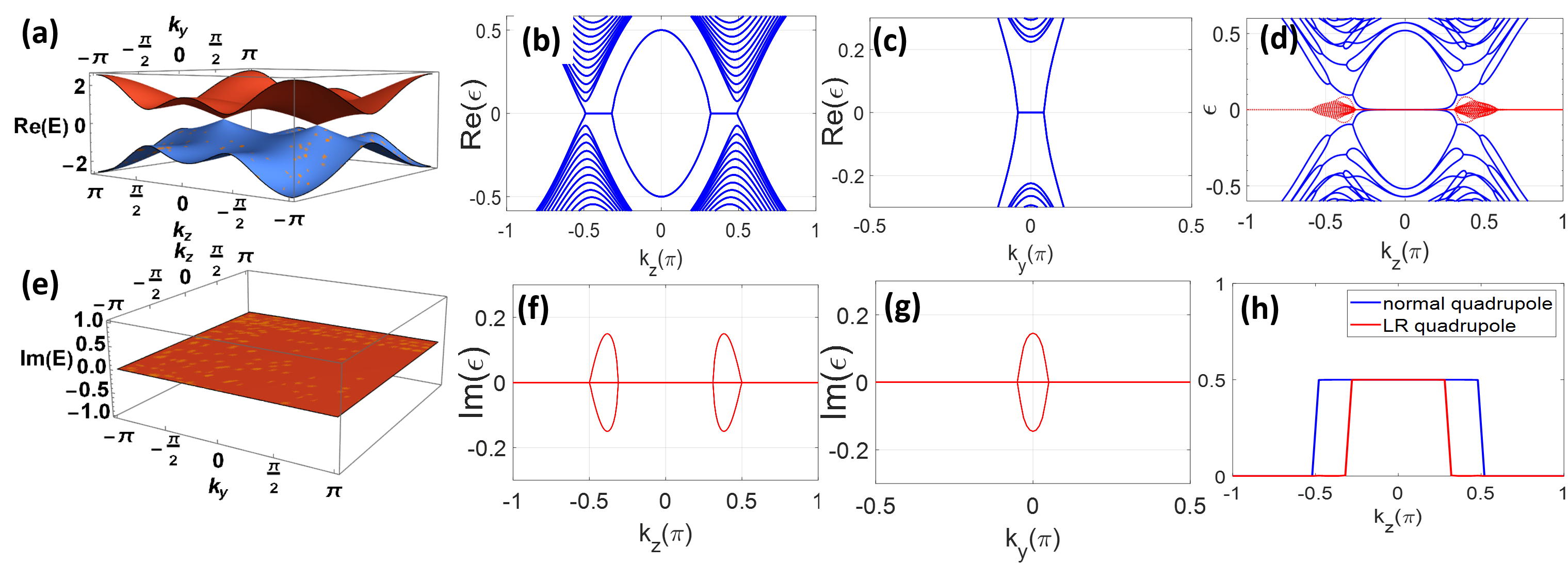}
    \caption{The spectrum of $\mathcal{T}$-NHHODSM with $\gamma=-1,\, m=-0.85$. The real and imaginary parts of the bulk spectrum in the $k_y-k_z$-plane are shown in (a,e). Real and imaginary spectra for open boundaries in $x$ are shown for  (b,f) the $k_z$-axis, (c,g) the line  $(k_y,\,k_z=0.45\pi)$. (d) the real (blue) and imaginary (red) spectra for open boundaries in $x$ and $y$ along $k_z$. (h) The $q^{RR}_{xy}$ (blue) and $q^{LR}_{xy}$ vs. $k_z$.}
    \label{fig:T-NHHODSM}
\end{figure*}
For the remainder of this article we focus on the regime of $m_1< m_1^c$ where the bulk spectrum is fully real. Remarkably,  for some $ m_1 < m_1^c$, the real HAs are destroyed and the surface develops EFRs formed between the projection of the two Dirac nodes (Fig.~\ref{fig:I-NHHODSM}).  Unlike previous work\cite{Edge-gainNatcomm2020,Edge-gainPRL2020,SatoPRX2019}, the surface EFRs do not emanate from an existing real state, since the surface in the Hermitian limit, $m_1=0$, is gapped. Instead, they suddenly appear in the gapped region between projections of the two bulk Dirac nodes simultaneously with the gapping out of the HAs. As mentioned above, when further increasing $m_1$ to $m_1>m_1^c$, the bulk spectrum becomes complex and the projections of the bulk ESs on the surfaces are shown in \cite{sm}.\\

Now let us discuss the topological protection of the surface EFRs in Fig.~\ref{fig:I-NHHODSM}(c-f). On the $x$-surface, the $H^\mathcal{I}_D$ in \eqref{I-NHHODSM} hosts $\mathcal{R}$, $\mathcal{T}\mathcal{M}_y$, $\mathcal{M}_z$, $\Gamma_0$ symmetries. However, only the product $\mathcal{T} \mathcal{M}_y \mathcal{M}_z$ and $\Gamma_0$ keep the surface momentum ($k_y$, $k_z$)
invariant, and thus may protect non-Hermitian gapless structures at an arbitrary position in momentum space. Since $(\mathcal{T} \mathcal{M}_y \mathcal{M}_z)^2 = 1$ and
$\{ \mathcal{T} \mathcal{M}_y \mathcal{M}_z, \Gamma \} = 0$, these symmetries define class $\mathcal{P}$CI\cite{NHclassSatoPRL2019}. The surface possesses a point gap, and since the surface EFRs shown in Fig.~\ref{fig:I-NHHODSM}(c-f)[also Fig.~\ref{fig:adpic}(a)] are co-dimension $1$ exceptional rings, the EFRs can be characterized by a $\mathbb{Z}$ topological invariant\cite{note1} which protects the degeneracies. We note that the bulk and surface having ESs protected by different symmetries is remarkable because it again shows the complicated bulk-boundary correspondence for non-Hermitian systems. More strikingly, since the bulk and surface are anti-PT and PT symmetric, respectively, both PT and anti-PT applications can be exploited in a single system in a natural way\cite{anti-PTPRAoriginal}. We call this new family of non-Hermitian topological phases, the \emph{hybrid-PT topological phases}.\\
\emph{$\mathcal{T}$-Model}.---Next we will consider perturbing Eq. \ref{hodsm1} with a non-Hermtian term that preserves $\mathcal{T}$:
\begin{align}\label{T-NHHODSM}
    H^{\mathcal{T}}_D = H_{HODSM}(\vex{k}) + im_2\Gamma_0 a_0(\vex{k})\sin(k_z).
\end{align}
Similar to $H^{\mathcal{I}}_D$, Eq.~\eqref{fig:T-NHHODSM} belongs to class $\mathcal{P}$CII$^{\dagger}$ and possesses  anti-PT symmetry. As before, the bulk spectrum of \eqref{T-NHHODSM} becomes complex only when $m_2 > m_2^c={\rm{min}}_{\bf{k}_0}\sum^4_i a^2_i(\vex{k}_0)/(a_0(\vex{k}_0)\sin(k_{z0}))^2$ \cite{note2}.  Similar to the case of $H^{\mathcal{I}}$, as $m_2$ is increased from zero, the surface spectrum (on surfaces normal to $\hat{x}$ or $\hat{y}$) becomes complex before the bulk. However, in this regime, unlike the $\mathcal{I}$-symmetric model, here the two Dirac nodes are connected by two EFRs which are separated by a real gap in the middle of the spectrum (Fig.~\ref{fig:T-NHHODSM}(b)). Remarkably, by further opening boundary along the $x$ and $y$-directions to get a hinge, we find Fermi hinge-arcs that survive in a region corresponding to the gapped region of the surface states (Fig.~\ref{fig:T-NHHODSM}(d)). Interestingly, the hinge arcs connected the two EFRs on the surface instead of bulk Dirac nodes.
Thus, the model in \eqref{T-NHHODSM}, is a type of NHHODSM which we call \emph{hybrid-order exceptional Dirac semimetal} since it has surface EFRs and hinge states.
As for symmetries, the surface EFRs on a surface perpendicular to the $x$-direction, $\mathcal{T} \mathcal{M}_y\mathcal{M}_z$ and $\Gamma_0$ symmetries are preserved. Thus, we obtain the same topological classification class  $\mathcal{P}$CI for the surface EFRs like the $\mathcal{I}$-symmetric case. Therefore, we find that the $H^{\mathcal{T}}_D$ like $H^{\mathcal{I}}_D$ is a hybrid-order PT topological phase exhibit anti-PT and PT symmetries in the bulk and surfaces, respectively.

\emph{Topological invariant}.--- To characterize the higher-order topology in the presence of non-Hermitian perturbations, we will adapt the calculation of the quadrupole moment $q_{xy}$ a non-Hermitian context. To do so, we employ the real-space, operator-based, formula  \cite{Qxyoperator1,Qxyoperator2,Qxyoperator3}. This bulk characteristic is crucial because of the complications due to the non-Hermitian bulk-boundary correspondence, e.g.,  explicitly calculating the corner charge and surface polarization (and hence $q_{xy}$) in non-Hermitian systems may be challenging due to ESs or the skin-effect. Let us now show that this formalism correctly captures the existence of zero-energy modes on the hinges in both the $\mathcal{I}$ and $\mathcal{T}$-symmetric models, provided we use a biorthogonal basis \cite{BiorthogonalPRL2018}:
\begin{align}
    q_{xy}=\frac{1}{2\pi}\im\bigg[\ln{\langle{\Psi^{R(L)}|\hat{U}_{xy}|\Psi^{R(L)}\rangle}}\bigg]
\end{align}
where $\hat{U}_{xy}=e^{2 \pi i\sum_r r_x r_y \hat{n}_r/(L_x L_y)},\,|\Psi^{R(L)}\rangle = \prod_{n \in occ} \gamma^{\dagger}_{n,R(L)}|0\rangle$, $\Psi^{R(L)}$ denote the right and left eigenvectors, and $r_x$ ($L_x$) and $r_y$ ($L_y$) are the coordinates (sample size) along the $x$ and $y$ directions, respectively.
Figure ~\ref{fig:T-NHHODSM}(h), shows the $q_{xy}$ for the model in Eq.~\eqref{T-NHHODSM} calculated using both normal ($RR$) and biorthogonal, $LR$, approaches. Interestingly, the $LR$ approach returns the correct value, i.e, it depicts quantized $q_{xy}=0.5$ for the region of momentum space in which there exist gapless hinge arcs, in the regime when the surface has EFRs, but the bulk is still fully real. Additionally, for  $H^{\mathcal{I}}_D$, we find the $q^{LR}_{xy}=0$ for all $k_z.$ In comparison, for both $H^{\mathcal{T}}_D$ and $H^{\mathcal{I}}_D$ $q^{RR}_{xy}$ returns $0.5$ for any $k_z$ between the bulk Dirac nodes, even for $k_z$ where there are no hinge-arcs such as the $H^{\mathcal{I}}_D$ case \cite{sm}.
\begin{figure}[tb!]
    \centering
    \includegraphics[width=0.45\textwidth]{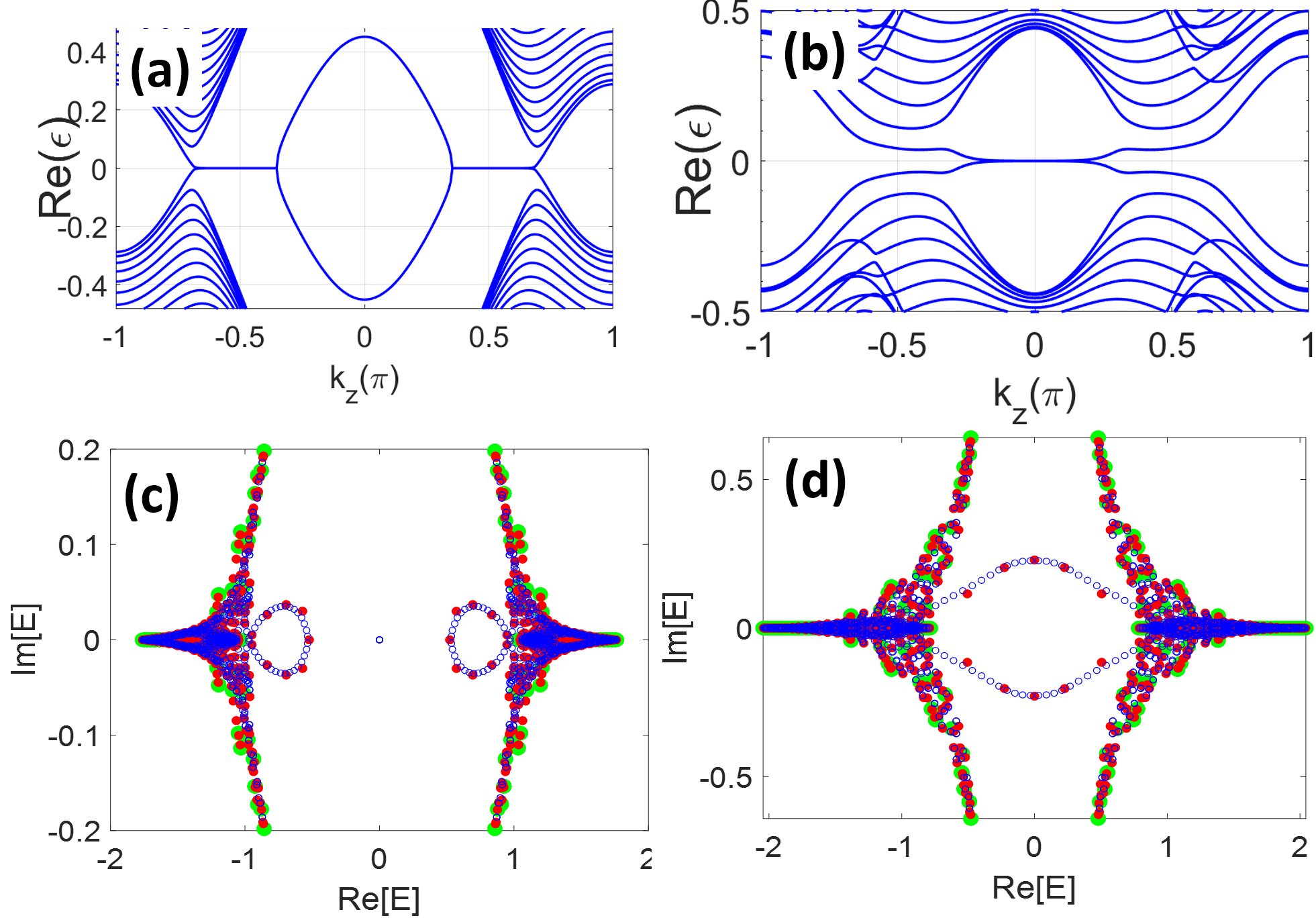}
    \caption{The real part of the spectrum of for Eq.~\eqref{I-NHHODSM} having $\gamma=-0.7,\,m_1=0.75,\,\alpha_2=0.1$ along (a) the $k_z$-axis for open-boundaries in the $x$-direction and (b) similar to (a) but with open boundaries along $x$ and $y$ to show hinge states. Next we show the complex-spectrum for $H^{\mathcal{T}}_D$ with the $\alpha_2=0.1$, for (c) $k_z=0.25\pi$ (i.e., hinge states present) and (d) $k_z=0.4\pi$ (i.e., hinge states absent). The filled green and red dots denote full periodic boundary conditions and $x$-open/$y$-periodic, respectively, while blue circles show the case when both $x,y$ are open boundaries. $L_x=20,\,L_y=20$ for all the boundary conditions in (c,d).}
    \label{fig:ExPert}
\end{figure}

\emph{Alternative Surface Phenomena}.--- By breaking some of the protective non-Hermitian symmetries of the surface we can generate other types of hybrid surface/hinge phenomena.  To show this, let us investigate the effect of Hermitian perturbations which leave the bulk Dirac nodes intact, and preserve the bulk $C^z_4$ symmetry, while breaking the surface protective non-Hermitian symmetries. The first example is $\alpha_1\Gamma_0 a_0(\vex{k})$ ($\alpha_1\Gamma_0 a_0(\vex{k})\sin(k_z)$), with strength $\alpha_1$ for the $H^{\mathcal{I}}_D$ ($H^{\mathcal{T}}_D$) model. In the case of $H^{\mathcal{I}}_D$ any bulk ESs (i.e., if $m_1>m_1^c$) become gapped when $\alpha_1\neq 0,$ while the real Dirac nodes remain gapless. The surface EFRs become gapped as well, however, since the perturbation gaps out the $x$ and $y$ surfaces with opposite signs, a non-zero $\alpha_1$ induces hinge domain walls (Fig.~\ref{fig:adpic}(a1)). Interestingly, the resulting phase is a $2nd$-order $\mathcal{I}$-NHHODSM having two Dirac nodes in the bulk and hinge states despite having no hinge modes when $\alpha_1=0$. Next, for $H^{\mathcal{T}}_D$ with $\alpha_1\neq 0,$ the bulk Dirac nodes remain gapless, and any bulk ESs are gapped. Moreover, the surface EFRs also become gapped while the hinge states remain intact (Fig.~\ref{fig:adpic}(b1)). Interestingly, even though the surface EFRs are gapped, the hinges remain arc-like and their lengths remain same as the case of $\alpha_1=0$ and so the systems is still a $\mathcal{T}$-NHHODSM.

To illustrate a second set of behaviors we consider adding the term $\alpha_2\sigma^3\kappa^2 a_0(\vex{k})$($\alpha_2\sigma^3\kappa^2 a_0(\vex{k})\sin(k_z)$) for $H^{\mathcal{I}}_D$ ($H^{\mathcal{T}}_D$). Like the previous case, $\alpha_2 \neq 0$ gaps out the bulk ESs for both $H^{\mathcal{I}}_D$, and $H^{\mathcal{T}}_D$ while preserving the real, bulk Dirac nodes. However, these terms do not fully gap out the surface EFRs, but instead deform them to EPs forming what we call an \emph{exceptional Fermi-arc}, i.e, a surface Fermi-arc terminated by an EPs instead of a conventional nodal point (Fig.~\ref{fig:adpic}(a2,b2)). Additionally, for the case of open boundary conditions in both the $x$ and $y$-directions we find the following remarkable results: (i) considering $H^{\mathcal{I}}_D$ having $\alpha_2 \neq 0$, we find that in the middle of the exceptional surface Fermi-arcs a gap opens up, as $\gamma$ is tuned, and hinge-arcs appear in the surface gap (Fig.~\ref{fig:ExPert}(b)) characterized by a quantized $q^{LR}_{xy}$ (see \cite{sm}). We recall that $H^{\mathcal{I}}_D$ with $\alpha_2=0$ does not possess hinge-arcs, therefore, $H^\mathcal{I}_D$ with non-zero $\alpha_2$ becomes an \emph{intrinsic hybrid-order non-Hermitian Dirac semimetal}.  (ii) Considering $H^{\mathcal{T}}_D$ and non-zero $\alpha_2$, we find that the HAs remain intact (having $q^{LR}_{xy}=0.5$) since $\alpha_2$ preserves $C^z_4$. Interestingly, the surface spectra, in the same region of $k_z$-space that the real HAs exist, develops complex loop spectra. For example, Figure.~\ref{fig:ExPert}(c,d) shows the complex spectra under various boundary conditions for a fixed-$k_z$ slice in the region of momentum space with and without HAs. Interestingly, for each value of $k_z$ in the region exhibiting HAs we find the formation of two loops having a point-gap with non-vanishing winding in the complex energy spectra (see Fig.~\ref{fig:ExPert}(c)), which disappear as $k_z$ is tuned and the two loops close/merge along imaginary axis (see Fig.~\ref{fig:ExPert}(d)). The result is that for each value of $k_z$ for which the real HAs exist there are also $\mathcal{O}(L)$ corner skin modes that appear which arise from the complex surface states\cite{2ndorderskinSatoPRB2020,2ndorderskinOkugawaPRB2020,2ndorderskinFuPRB2021}. Therefore, we find a novel $3d$ non-Hermitian topological semimetal having coexisting real HAs and hinge skin modes (HSMs), which we call a \emph{higher-order skin-topological Dirac semimetal}.

\emph{Concluding remarks}.---We have investigated higher-order Dirac semimetals in the presence of $C^z_4$-symmetric non-Hermitian perturbations. We have shown that they exhibit the striking property that bulk and surfaces are anti-PT and PT symmetric, respectively, realizing a novel non-Hermitian topological phase which we dubbed, a hybrid-PT topological phase. These systems may open up new avenues for many fundamental and practical directions, such as novel topological lasers, sensors and photonics \cite{anti-PTPRAoriginal,feng2017non, peng2016anti,PhysRevLett.113.123004,antonosyan2015parity} with selected utilization of anti-PT/PT on the bulk/surface. In our models in the bulk, real Dirac nodes and ESs coexist, providing a platform for utilizing both Dirac and exceptional physics in the bulk as well as surfaces in the same system. Moreover, we revealed that they show a unique edge-gain (complex surface) without bulk-gain (real bulk). \\
Finally we make brief remarks on plausible experimental directions. The main building block of models discussed here, i.e., stacks of 2D quadrupole insulators have been realized in variety of platforms both in Hermitian \cite{Peterson2018,Noh2018,Serra-Garcia2018,Imhof2018,ni2019observation,xue2019acoustic} and non-Hermitian \cite{zhang2ndordersonic} systems. Recently, 3D models of Hermitian HOSMs have been implemented in experiments \cite{wei2021higher,luo2021observation,ni2021higher,qiu2020higherorder}. Therefore, the physics discussed in this work is readily accessible and can motivate an immediate experimental realization of NHHODSMs.


\emph{Acknowledgement}.---S.A.A.G acknowledges support
from ARO (Grant No. W911NF-18-1-0290) and NSF
(Grant No. DMR1455233). T.L. and T.L.H. thank the US Office of Naval Research (ONR) Multidisciplinary University Research Initiative (MURI) grant N00014-20-1-2325 on Robust Photonic Materials. M.S. was supported by JST CREST Grant No. JPMJCR19T2, Japan, and KAKENHI Grant No. JP20H00131 from the JSPS.

\bibliography{NHhodsm}

\begin{thebibliography}{70}
\bibitem [{not({\natexlab{c}})}]{note2}%
  \BibitemOpen
  \href@noop {} {\bibfield  {journal} {\bibinfo  {journal} {Note that due to
  explicit $k_z$ dependence, the non-Hermitian term vanishes at high-symmetry
  points of $k_z$. Therefore, a avoided region appear in the spectrum and forms
  two patches of ESs on opposite momenta in the vicinity of $k_z=0,\pi$ of bulk
  spectrum.}\ } ({\natexlab{c}})}\BibitemShut {NoStop}%
\end{thebibliography}

\end{document}